# Quantum Multiple Access Wiretap Channel: On the One-Shot Achievable Secrecy Rate Regions

Hadi Aghaee
Faculty of Electrical Engineering
K. N. Toosi University of Technology
Tehran, Iran
Email: Aghaee_Hadi@email.kntu.ac.ir

Bahareh Akhbari
Faculty of Electrical Engineering
K. N. Toosi University of Technology
Tehran, Iran
Email: akhbari@eetd.kntu.ac.ir

*Abstract—* In this paper, we want to investigate classical-quantum multiple access wiretap channels (CQ-MA-WTC) under one-shot setting. In this regard, we analyze the CQ-MA-WTC using simultaneous position-based decoder for reliable decoding and using a newly introduced technique in order to decode securely. Also, for the sake of comparison, we analyze the CQ-MA-WTC using Sen's one-shot joint typicality lemma for reliable decoding. The simultaneous position-based decoder tends to a multiple hypothesis testing problem. Also, using convex splitting to analyze the privacy criteria in a simultaneous scenario becomes problematic. To overcome both problems, we first introduce a new channel that can be considered as a dual to the CQ-MA-WTC. This channel is called a point-to-point quantum wiretap channel with multiple messages (PP-QWTC). In the following, as a strategy to solve the problem, we also investigate and analyze quantum broadcast channels (QBCs) under the one-shot setting.

*Keywords—Quantum Channel; Mutual Information; Secrecy Capacity; Multiple Access Channel*

## I. INTRODUCTION

The quantum multiple access channel (QMAC) was first introduced by Winter [1]. A QMAC can accept two or more messages (classical or quantum) as inputs and one output.

Similar to the classical world, decoding messages over a QMAC is based on two main techniques: successive cancelation decoding and simultaneous decoding. In [1], the author employs the successive cancelation decoding technique.

A quantum broadcast channel (QBC) is a channel with a sender and two or more receivers. The sender wishes to transmit two or more messages (classical or quantum) over the channel to the receivers. The QBC was first introduced by Yard *et al.* [2]. In [2], the authors derived an inner bound for QBC for i.i.d. (independent and identical) case, and in [3], the authors derived the same inner bound using a more straightforward method and more in the spirit of its classical analogous [4] than the method in [2].

In recent decades, with development of quantum data processing and its applications, the necessity to study the security of quantum channels has increased. In this regard, the quantum wiretap channel (QWTC) was first introduced in [5] and [6].

Then, the secrecy constraints are extended to multi-user quantum channels such as quantum interference channel (QIC) [7,8], and quantum multiple access channel (QMAC) [9-13].

There are two bottlenecks in studying the security of quantum channels. The first is decoding three or more messages simultaneously (reliability), and the second is about how we can securely decode two or more messages (confidentiality). The first bottleneck arises from the nonexistence of a general quantum joint typicality lemma. However, this problem has been solved in some cases, such as the min-entropy case and QMACs with commutative output [14]. Therefore, in the independent and identical distributed (i.i.d.) case, successive decoding combined with time-sharing techniques should be used. In this setting, transmitters are allowed to transmit their messages by only one use channel. Sen proved a joint typicality lemma which helps to decode any number of messages simultaneously in the one-shot case [14]. Obtaining secrecy against the eavesdropper by Wyner's technique [15] of randomizing over a block becomes problematic in the quantum setting. Wyner's technique has been shown to work for point-to-point quantum channels by Devetak [6] and explained further in [16]. However, there are no easy generalizations to multiple senders for a quantum channel. This issue is discussed in detail in [16].

In this paper, we want to investigate the secrecy problem of quantum multiple access channel (QMAC) with classical inputs under one-shot setting. Also, we have investigated some bottlenecks connected to decoding process for CQ-MA-WTC. The achievement of this paper is about analyzing bottlenecks in decoding process and providing solutions to overcome them.

Also, we present two techniques for quantum multiple access wiretap channel with classical inputs (CQMA-WTC). The first approach is based on the method presented in [14], and another technique is the *simultaneous position-based decoder*. From [17], we know that the simultaneous position-based decoder tends to a multiple quantum hypothesis testing problem which is solvable in a special case. Also, from [18], we know that the convex split lemma could not be used to analyze the privacy of multiple messages in simultaneous decoding.

The paper is organized as follows: In Section II, some seminal definitions are presented. In Section III, the main channel and information processing tasks are presented. In Section IV, the results and main theorems are presented. Section V is dedicated to discussion.

## II. PRELIMINARIES

Let A (Alice), B (Bob), and C (Charlie) be three quantum systems. These quantum systems can be denoted by their corresponding Hilbert spaces as $\mathcal{H}^A$, $\mathcal{H}^B$, and $\mathcal{H}^C$. The states of the above quantum systems are presented as density operators $\rho^A$, $\rho^B$, and $\rho^C$, respectively, while the shared state between Alice, Bob, and Charlie is denoted by $\rho^{ABC}$. A density operator is a positive semidefinite operator with a unit trace. Alice, Bob, or Charlie's state can be defined by a partial trace operator over the shared state. The partial trace is used to model the lack of access to a quantum system. Thus, Alice's density operator using partial trace is $\rho^A = Tr_{BC}\{\rho^{ABC}\}$. $|\psi\rangle^A$ denotes the pure state of system A. The corresponding density operator is $\psi^A = |\psi\rangle\langle\psi|^A$. The von Neumann entropy of the state $\rho^A$ is defined by $H(A)_\rho = -Tr\{\rho^A \log \rho^A\}$. For an arbitrarily state such as $\sigma^{AB}$, the quantum conditional entropy is defined by $H(A|B)_\sigma = H(A,B)_\sigma - H(B)_\sigma$. The quantum mutual information is defined by $I(A;B)_\sigma = H(A)_\sigma + H(B)_\sigma - H(A,B)_\sigma$, and the conditional quantum mutual information is defined by:

$$I(A;B|C)_\sigma = H(A|C)_\sigma + H(B|C)_\sigma - H(A,B|C)_\sigma$$

Quantum operations can be denoted by *completely positive trace-preserving* (CPTP) maps $\mathcal{N}^{A\to B}$. The CPTP maps accept input states in A and output states in B. The distance between two quantum states, such as A and B, is defined by trace distance. The trace distance between two arbitrary states, such as $\sigma$ and $\rho$ is:

$$\|\sigma - \rho\|_1 = Tr|\sigma - \rho| \qquad (1)$$

where $|\Psi| = \sqrt{\Psi^\dagger \Psi}$. This quantity is zero for two similar and perfectly distinguishable states.

*Fidelity* is defined as $F(\rho,\sigma) = \left\|\sqrt{\rho}\sqrt{\sigma}\right\|_1^2$, and *purified distance* is a metric on $\mathcal{D}(\mathcal{H})$ and is defined as $P(\rho,\sigma) := \sqrt{1 - F(\rho,\sigma)^2}$. Most of the above definitions are given in [19].

**Definition 1:** *(Hypothesis testing mutual information)* Hypothesis testing mutual information is denoted by $I_H^\epsilon(X;Y) := D_H^\epsilon(\rho_{XY} \| \rho_X \otimes \rho_Y), \epsilon \in (0,1)$ and is considered as *quantum hypothesis testing divergence* [17] where $D_H^\epsilon(.\|.)$ is *hypothesis testing relative entropy* [17]. $\epsilon$ is the smoothing variable, $\rho^{\mathcal{H}_X \mathcal{H}_Y}$ is the joint classical-quantum state of input and output over their Hilbert spaces $(\mathcal{H}_X, \mathcal{H}_Y)$, and it can be shown as $\rho_{XY}$:

$$\rho_{XY} = \sum_x p_X(x)|x\rangle\langle x|_X \otimes \rho_Y^x$$

where $p_X$ is the input distribution.

**Definition 2:** *(Quantum relative entropy [20]):* Consider states $\rho_X, \sigma_X \in \mathcal{D}(\mathcal{H}_X)$. The Quantum relative entropy is defined as:

$$D(\rho_X \| \sigma_X) := \begin{cases} Tr\{\rho_X[\log_2 \rho_X - \log_2 \sigma_X]\} & supp(\rho_X) \subseteq supp(\sigma_X) \\ +\infty & otherwise \end{cases}$$

where $supp(\sigma_X)$ refers to the *set-theoretic support* of $\sigma$. $supp(\sigma)$ is the subspace of $\mathcal{H}$ spanned by all eigenvectors of $\sigma$ with non-zero eigenvalues.

**Fact 1**: The following relation exists between the quantum relative entropy and hypothesis testing relative entropy for $\epsilon \in (0,1)$ [21]:

$$D_H^\epsilon(\rho_X \| \sigma_X) \le \frac{1}{1-\epsilon}[D(\rho_X \| \sigma_X) + h_b(\epsilon)]$$

where $h_b(\epsilon) := -\epsilon \log_2 \epsilon - (1-\epsilon)\log_2(1-\epsilon)$ is a binary entropy function.

**Definition 3**: *(Max mutual information [21])* Consider a bipartite state $\rho_{XY}$ and a parameter $\epsilon \in (0,1)$. The max mutual information can be defined as follows:

$$I_{max}(X;Y)_\rho := D_{max}(\rho_{XY} \| \rho_X \otimes \rho_Y)_\rho$$

where $\rho$ refers to the state $\rho_{XY}$ and $D_{max}(\|)$ is the *max-relative entropy* [22] for $\rho_X, \sigma_X \in \mathcal{H}_X$:

$$D_{max}(\rho_X \| \sigma_X) := \inf\{\gamma \in \mathbb{R}: \rho_X \le 2^\gamma \sigma_X\}$$

**Definition 4**: *(Quantum smooth max relative entropy [22])* Consider states $\rho_X, \sigma_X \in \mathcal{D}(\mathcal{H}_X)$, and $\epsilon \in (0,1)$. The quantum smooth max relative entropy is defined as:

$$D_{max}^\epsilon(\rho_X \| \sigma_X) := \inf_{\rho_X' \in \mathcal{B}^\epsilon(\rho_X)} D_{max}(\rho_X' \| \sigma_X)$$

where $\mathcal{B}^\epsilon(\rho_X) := \{\rho_X' \in \mathcal{D}(\mathcal{H}_X): P(\rho_X', \rho_X) \le \epsilon\}$ is $\epsilon$-ball for $\rho_{XY}$.

**Definition 5**: *(Quantum smooth max mutual information [21])* Consider $\rho_{XY} := \sum_{x \in \mathcal{X}} P_X(x)|x\rangle\langle x|_X \otimes \rho_Y^x$ as a classical-quantum state and a parameter $\epsilon \in (0,1)$. The smooth max mutual information between the systems $X$ and $Y$ can be defined as follows:

$$I_{max}^\epsilon(X;Y) := \inf_{\rho_{XY}' \in \mathcal{B}^\epsilon(\rho_{XY})} D_{max}(\rho_{XY}' \| \rho_X \otimes \rho_Y)$$
$$= \inf_{\rho_{XY}' \in \mathcal{B}^\epsilon(\rho_{XY})} I_{max}(X;Y)_{\rho'},$$

where $\mathcal{B}^\epsilon(\rho_{XY}) := \{\rho_{XY}' \in \mathcal{D}(\mathcal{H}_X \otimes \mathcal{H}_Y): P(\rho_{XY}', \rho_{XY}) \le \epsilon\}$ is $\epsilon$-ball for $\rho_{XY}$.

**Definition 6**: *(Conditional smooth hypothesis testing mutual information [23])* Consider $\rho_{XYZ} := \sum_{z \in Z} p_Z(z)|z\rangle\langle z|_Z \otimes \rho_{XY}^z$ be a tripartite classical-quantum state and $\epsilon \in (0,1)$. We define,

$$I_H^\epsilon(X;Y|Z)_\rho := \max_{\rho'} \min_{z \in supp(\rho_Z')} I_H^\epsilon(X;Y)_{\rho_{XY}^z},$$

where maximization is over all $\rho_Z' = \sum_{z \in Z} p_Z(z)|z\rangle\langle z|_Z$ satisfying $P(\rho_Z', \rho_Z) \le \epsilon$.

**Fact 2**: [24] Let $\rho_{XYZ} := \sum_{z \in Z} p_Z(z)|z\rangle\langle z|_Z \otimes \rho_{XY}^z$ be a tripartite classical-quantum state and $\epsilon \in (0,1)$, the following relation holds,

$$\lim_{n\to\infty} \frac{1}{n} I_H^\epsilon(X^{\otimes n}; Y^{\otimes n} | Z^n)_{\rho^{\otimes n}} = I(X;Y|Z)_\rho$$

**Definition 7**: *(Alternate smooth max-mutual information)* Consider a bipartite state $\rho_{XY}$ and a parameter $\epsilon \in (0,1)$. The *alternate* definition of the *smooth max-mutual information* between the systems $X$ and $Y$ can be defined as follows:

$$\tilde{I}_{max}^\epsilon(Y;X) := \inf_{\rho_{XY}' \in \mathcal{B}^\epsilon(\rho_{XY})} D_{max}(\rho_{XY}' \| \rho_X \otimes \rho_Y')$$

**Fact 3**: (Relation between two definitions of the smooth max mutual information) [25]: Let $\epsilon \in (0,1)$ and $\gamma \in (0,\epsilon)$ For a bipartite state $\rho_{XY}$, it holds that:

$$\tilde{I}^\epsilon_{max}(Y;X)_\rho \leq I^{\epsilon-\gamma}_{max}(X;Y)_\rho + \log\frac{3}{\gamma^2}$$

**Definition 8**: *(Conditional smooth max mutual information [23])* Consider $\rho_{XYZ} := \sum_{z \in Z} p_Z(z)|z\rangle\langle z|_Z \otimes \rho^z_{XY}$ be a tripartite classical-quantum state and $\epsilon \in (0,1)$. We define,

$$I^\epsilon_{max}(X;Y|Z)_\rho := \max_{\rho'} \min_{z \in supp(\rho'_Z)} I^\epsilon_{max}(X;Y)_{\rho^z_{XY}},$$

where maximization is over all $\rho'_Z = \sum_{z \in Z} p_Z(z)|z\rangle\langle z|_Z$ satisfying $P(\rho'_Z, \rho_Z) \leq \epsilon$.

**Fact 4**: [24] $\rho_{XYZ} := \sum_{z \in Z} p_Z(z)|z\rangle\langle z|_Z \otimes \rho^z_{XY}$ be a tripartite classical-quantum state and $\epsilon \in (0,1)$, the following relation holds,

$$\lim_{n \to \infty} \frac{1}{n} I^\epsilon_{max}(X^{\otimes n}; Y^{\otimes n}|Z^n)_{\rho^{\otimes n}} = I(X;Y|Z)_\rho$$

**Definition 9:** *(Quantum Rényi relative entropy of order $\alpha$ [17])* For a state $\rho \in \mathcal{D}(\mathcal{H})$ and a positive semidefinite operator $\sigma$, the *quantum Rényi relative entropy of order $\alpha$*, where $\alpha \in [0,1) \cup (1, +\infty)$ is defined as:

$$D_\alpha(\rho \| \sigma) \equiv \frac{1}{\alpha - 1} \log_2 Tr\{\rho^\alpha \sigma^{1-\alpha}\}$$

Also, *Rényi entropy of order $\alpha$* can be defined as follows:

$$H_\alpha(A)_\rho \equiv \frac{1}{1-\alpha} \log_2 Tr\{\rho^\alpha_A\}$$

**Definition 10**: *(One-shot inner bound of a classical-quantum multiple access channel)* [14] A two-user classical-quantum multiple access channel (C-QMAC) under the one-shot setting is a triple $(\mathcal{X}_1 \times \mathcal{X}_2, \mathcal{N}_{X_1 X_2 \to Y}(x_1, x_2) \equiv \rho^Y_{x_1 x_2}, \mathcal{H}_Y)$, where $\mathcal{X}_1$ and $\mathcal{X}_2$ are the alphabet sets of two classical inputs, and $Y$ is the output system. $\rho^Y_{x_1 x_2}$ is a quantum state, and the channel has a completely positive trace-preserving map (CPTP) $\mathcal{N}_{X_1 X_2 \to Y}$.

Considering the joint typicality lemma introduced in [Corollary 4, 14], the one-shot inner bound of a C-QMAC is as follows:

$$R_1 \leq I^\epsilon_H(X_1:X_2 Y)_\rho - 2 + \log \epsilon$$

$$R_2 \leq I^\epsilon_H(X_2:X_1 Y)_\rho - 2 + \log \epsilon$$

$$R_1 + R_2 \leq I^\epsilon_H(X_1 X_2:Y)_\rho - 2 + \log \epsilon$$

with decoding error at most $49\sqrt{\epsilon}$, where $I^\epsilon_H(.)$ is the hypothesis testing mutual information defined in Definition 1 with respect to the controlling state:

$$\rho^{Q X_1 X_2 Y} := \sum_{q x_1 x_2} p(q) p(x_1|q) p(x_2|q) |q x_1 x_2\rangle\langle q x_1 x_2|^{Q X_1 X_2} \otimes \rho^Y_{x_1 x_2} \quad (2)$$

and $Q$ is a time-sharing variable.

Note that $I^\epsilon_H(:)$ is the difference between a *Rényi entropy* of order two and a conditional quantum entropy.

**Lemma 1**: [16] *Given the control state in* (2) *(without time-sharing variable), $\delta' > 0$ and $0 < \epsilon' < \delta'$, let $\{x_1, \ldots, x_{K_1}\}$ and $\{y_1, \ldots, y_{K_2}\}$ be i.i.d. samples from the distributions $P_X$ and $P_Y$. Then, if*

$$\log|\mathcal{K}_1| \geq I^{\delta'-\epsilon'}_{max}(X:Z)_\rho + \log\frac{3}{\epsilon'^3} - \frac{1}{4}\log \delta'$$

$$\log|\mathcal{K}_2| \geq I^{\delta'-\epsilon'}_{max}(Y:ZX)_\rho + \log\frac{3}{\epsilon'^3} - \frac{1}{4}\log \delta' + \mathcal{O}(1)$$

*the following holds,*

$$\mathbb{E}_{\substack{x_1,\ldots,x_{K_1} \sim P_X \\ y_1,\ldots,y_{K_2} \sim P_Y}} \left\| \frac{1}{|\mathcal{K}_1||\mathcal{K}_2|} \sum_{j=1}^{|\mathcal{K}_2|} \sum_{i=1}^{|\mathcal{K}_1|} \rho^Z_{x_i y_j} - \rho^Z \right\|_1 \leq 20 \delta'^{\frac{1}{8}}$$

*Proof*: see [16].

**Lemma 2:** *(Convex split lemma)* [19,20] Let $\rho_{XY}$ be an arbitrary state and suppose that $\tau_{X_1 \ldots X_k B}$ be the following state:

$$\tau_{X_1 \ldots X_k B} = \frac{1}{K} \sum_{k=1}^{K} \rho_{X_1} \otimes \ldots \otimes \rho_{X_{k-1}} \otimes \rho_{X_k B} \otimes \rho_{X_{k+1}} \otimes \ldots \otimes \rho_{X_k}$$

Let $\epsilon \in (0,1)$ and $\delta \in (0, \sqrt{\epsilon}]$, if

$$\log_2 K = \tilde{I}^{\sqrt{\epsilon}-\delta}_{max}(Y;X)_\rho + 2 \log_2\left(\frac{1}{\delta}\right)$$

then,

$$P(\tau_{X_1 \ldots X_k B}, \rho_{X_1} \otimes \ldots \otimes \rho_{X_k} \otimes \tilde{\rho}_Y) \leq \sqrt{\epsilon}$$

for some state $\tilde{\rho}_Y$ such that $P(\rho_Y, \tilde{\rho}_Y) \leq \sqrt{\epsilon} - \delta$.

*Proof*: see [20].

**Lemma 3:** *(Hayashi-Nagaoka inequality [26])* Suppose that $S, T \in \mathcal{P}(\mathcal{H}_X)$ such that $(I - S) \in \mathcal{P}(\mathcal{H}_X)$ are operators such that $T \geq 0$ and $0 \leq S \leq I$, then for all positive constant $c$, the following relation holds:

$$I - (S + T)^{-\frac{1}{2}} S (S + T)^{-\frac{1}{2}} \leq (1 + c)(I - S) + (2 + c + c^{-1}) T$$

*Proof*: see [26].

### III. CHANNEL MODEL

A two-user CQ-MA-WTC is a triple $(\mathcal{X}_1 \times \mathcal{X}_2, \mathcal{N}^{X_1 X_2 \to YZ}(x_1, x_2) \equiv \rho^{YZ}_{x_1 x_2}, \mathcal{H}^Y \otimes \mathcal{H}^Z)$, where $\mathcal{X}_i, i \in \{1,2\}$ denote the input alphabet sets, and $Y, Z$ denote the output systems ($Y$ denotes the channel output at the legitimate receiver (Charlie), and $Z$ is the channel output at the eavesdropper). $\rho^{YZ}_{x_1 x_2}$ is the system output's quantum state. Both users want to transmit their messages as secure as possible over a CQ-MA-WTC to the receiver.

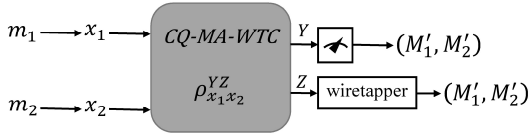

*Figure 1. The CQ-MA-WTC model*

The main channel is illustrated in Figure 1.

Consider the main channel illustrated in Figure 1. Each user chooses its message $m_i; i \in \{1,2\}$ from its message set $\mathcal{M}_i = [1:|\mathcal{M}_i| = 2^{R_i}]; i \in \{1,2\}$ ($R_1$ and $R_2$ are the transmitting rates corresponding to the first and the second messages, respectively), and sends it over a CQ-MA-WTC. The users also use two junk variables $k_i; i \in \{1,2\}$ from two amplification sets $\mathcal{K}_i = [1:|\mathcal{K}_i| = 2^{\hat{R}_i}]; i \in \{1,2\}$ for randomizing Eve's knowledge. We have two doubly indexed codebooks $x_1(m_1, k_1)$, and $x_2(m_2, k_2)$, for user-1 and user-2, respectively.

## IV. MAIN RESULTS

In this section, we present the main results.

Corollary 1 gives a one-shot achievable secrecy rate region for sending classical messages over a CQ-MA-WTC based on Sen's quantum joint typicality lemma [14]. The second theorem presents a novel approach to decode both messages over a CQ-MA-WTC reliably and confidentially (simultaneous position-based decoder). It should be noted that Corollary 1 and Theorem 1 use the same method to prove the security requirements. Also, we present a theorem that tries to overcome the bottlenecks connected to Theorem 1.

***Corollary 1:*** *(One-shot achievable rate region for CQ-MA-WTC)* Consider a two-user CQ-MA-WTC which accepts $X_1$ and $X_2$ as inputs and $Y$, and $Z$ as outputs. $\rho_{x_1 x_2}^{YZ}$ is the channel density operator. For any fixed $\epsilon \in (0,1), \epsilon' \in (0, \delta')$ and $\delta, \delta'$ such that $\delta' > 0$, the rate pair $\left(R_1, R_2, 49\sqrt{\epsilon} + 20\delta'^{\frac{1}{8}}\right)$ is achievable to satisfy the following inequalities:

$$R_1 \leq I_H^\epsilon(X_1:X_2Y|Q)_\rho - I_{max}^\eta(X_1:Z|Q)_\rho + \log \epsilon - 2 - \log \frac{3}{\epsilon'^3}$$
$$+ \frac{1}{4}\log \delta'$$

$$R_2 \leq I_H^\epsilon(X_2:X_1Y|Q)_\rho - I_{max}^\eta(X_2:ZX_1|Q)_\rho + \log \epsilon - 2$$
$$- \log \frac{3}{\epsilon'^3} + \frac{1}{4}\log \delta' + \mathcal{O}(1)$$

$$R_1 + R_2 \leq I_H^\epsilon(X_1X_2:Y|Q)_\rho - I_{max}^\eta(X_1:Z|Q)_\rho$$
$$- I_{max}^\eta(X_2:ZX_1|Q)_\rho + \log \epsilon - 2 - 2\log \frac{3}{\epsilon'^3}$$
$$+ \frac{1}{2}\log \delta' + \mathcal{O}(1)$$

where $\eta = \delta' - \epsilon'$ and the union is taken over input distribution $p_Q(q)p_{X_1|Q}(x_1|q)p_{X_2|Q}(x_2|q)$. $Q$ is the time-sharing random variable, and all of the mutual information quantities are taken with respect to the following state:

$$\rho^{QX_1X_2YZ} \equiv \sum_{q,x_1,x_2} p_Q p(q) p_{X_1|Q}(x_1|q) p_{X_2|Q}(x_2|q) |q\rangle\langle q|^Q$$
$$\otimes |x_1\rangle\langle x_1|^{X_1} \otimes |x_2\rangle\langle x_2|^{X_2}$$
$$\otimes \rho_{x_1 x_2}^{YZ} \quad (3)$$

*Proof:* See Appendix A.

*Sketch of proof:* The proof has two steps: 1- Reliable decoding based on Sen's quantum one-shot joint typicality (Definition 9). 2- Secure decoding based on Lemma 1.

***Theorem 1:*** *(one-shot lower bound for CQ-MA-WTC)* For any fixed $\epsilon \in (0,1), \epsilon' \in (0,1)$ and $\delta, \delta'$ such that $\delta \in (0, \epsilon)$, and $\delta' \in (0, \epsilon')$, there exists a one-shot code for the channel $\mathcal{N}_{X_1 X_2 \to YZ}$, if rate pair $\left(R_1, R_2, \epsilon + 2\delta + 20\delta'^{\frac{1}{8}}\right)$ satisfies the following bounds:

$$R_1 \leq I_H^\epsilon(X_1:X_2Y|Q)_\rho - I_{max}^\eta(X_1:Z|Q)_\rho - \log_2\left(\frac{4\epsilon}{\delta^2}\right)$$
$$- \log \frac{3}{\epsilon'^3} + \frac{1}{4}\log \delta'$$

$$R_2 \leq I_H^\epsilon(X_2:X_1Y|Q)_\rho - I_{max}^\eta(X_2:ZX_1|Q)_\rho - \log_2\left(\frac{4\epsilon}{\delta^2}\right)$$
$$- \log \frac{3}{\epsilon'^3} + \frac{1}{4}\log \delta' + \mathcal{O}(1)$$

$$R_1 + R_2 \leq I_H^\epsilon(X_1X_2:Y|Q)_\rho - I_{max}^\eta(X_1:Z|Q)_\rho$$
$$- I_{max}^\eta(X_2:ZX_1|Q)_\rho - \log_2\left(\frac{4\epsilon}{\delta^2}\right)$$
$$- 2\log \frac{3}{\epsilon'^3} + \frac{1}{2}\log \delta' + \mathcal{O}(1)$$

where $\eta = \delta' - \epsilon'$ and the union is taken over input distribution $p_Q(q)p_{X_1|Q}(x_1|q)p_{X_2|Q}(x_2|q)$. $Q$ is the time-sharing random variable, and all mutual information quantities are taken with respect to the state (3).

*Proof:* See Appendix B.

*Sketch of proof:* The proof has two steps: 1- Reliable decoding based on the *simultaneous position-based technique:* for simplicity of analysis, we merge reliability and confidentiality criteria into a single criterion [20]. 2- Secure decoding based on the Lemma 1.

**Remark 1:** It should be noted that, both of the above theorems tend to the same result if and only if $\delta = \epsilon$.

As mentioned before, the simultaneous position-based decoder tends to a multiple hypothesis testing problem which is unsolvable in the general case. Also, the convex split lemma (Lemma 2) does not make sense in the simultaneous decoding. Because it runs to the famous smoothing bottleneck of quantum information theory.

Now, consider the channel illustrated in Figure 2.

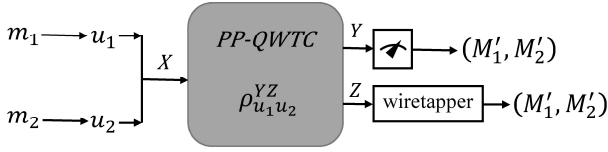

Figure 2. The PP-QWTC model.

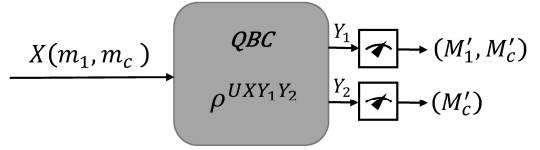

Figure 3. The QBC model.

This channel accepts two or more messages from one user. We call this channel a point-to-point quantum wiretap channel with multiple messages (PP-QWTC).

Consider PP-QWTC with classical messages. This channel is studied in [27] under a different scenario wherein a sender wants to send classical and quantum messages simultaneously to a legitimate receiver.

*Information processing task*: Two classical messages $(m_1, m_2) \in \mathcal{M}_1 \times \mathcal{M}_2$ are possessed by a sender (Alice) and be transmitted to a receiver (Bob) in the presence of a passive wiretapper over a point-to-point quantum channel under the one-shot scenario. Both of the messages, should be kept as secure as possible from the wiretapper. The PP-QWTC is a triple $(\mathcal{X}, \mathcal{N}^{\mathcal{X} \to YZ}(u_1, u_2) \equiv \rho_{x(u_1,u_2)}^{YZ}, \mathcal{H}^Y \otimes \mathcal{H}^Z)$, where $\mathcal{X}$ denotes the input alphabet sets, and $Y$, $Z$ denote the output systems ($Y$ denotes the channel output at the legitimate receiver (Bob), and $Z$ is the channel output at the eavesdropper). $\rho_{x(u_1,u_2)}^{YZ} \equiv \rho_{u_1 u_2}^{YZ}$ is the system output's quantum state.

Alice chooses its message $m_i$; $i \in \{1,2\}$ from its message set $\mathcal{M}_i = [1:|\mathcal{M}_i| = 2^{R_i}]$; $i \in \{1,2\}$, and sends it over a PP-QWTC. Alice also uses two junk variables $k_i$; $i \in \{1,2\}$ from two amplification sets $\mathcal{K}_i = [1:|\mathcal{K}_i| = 2^{\hat{R}_i}]$; $i \in \{1,2\}$ for randomizing Eve's knowledge. We have two doubly indexed codebooks $u_1(m_1, k_1)$, and $u_2(m_2, k_2)$.

*Encoding*: An encoding operation by Alice $\mathcal{E}: M_1 M_2 \to \mathcal{D}(\mathcal{H}_A)$

$$\forall m_1, m_2 \in M_1, M_2 \quad \frac{1}{2}\|\rho_{M_1 M_2 Z} - \rho_{M_1 M_2} \otimes \tilde{\rho}_Z\|_1 \leq \epsilon_2 \quad (4)$$

where for each message $m_1, m_2$, $\rho_{M_1 M_2}$ and $\rho_{M_1 M_2}$ are appropriate marginal of the state $\rho_{M_1 M_2 YZ} = \frac{1}{|\mathcal{M}_1||\mathcal{M}_2|}\sum_{m_2=1}^{|\mathcal{M}_2|}\sum_{m_1=1}^{|\mathcal{M}_1|}|m_1\rangle\langle m_1| \otimes |m_2\rangle\langle m_2| \otimes \mathcal{N}(\mathcal{E}(m_1, m_2))$. Also, $\tilde{\rho}_Z$ can be any arbitrary state.

*Decoding:* Decoding operation by Bob $\mathcal{D}: \mathcal{D}(\mathcal{H}_B) \to \widehat{M}_1 \widehat{M}_2$ such that:

$$pr\left(\left(\widehat{M}_1, \widehat{M}_2\right) \neq (M_1, M_2)\right) \leq \epsilon_1 \quad (5)$$

A rate pair $(R_1, R_2)$ is $(\epsilon_1, \epsilon_2)$-achievable if, for such encoding and decoding maps $(\mathcal{E}, \mathcal{D})$, the conditions stated in (4) and (5) are satisfied.

As it can be understood from criterion (4), the reliability and confidentiality conditions are merged into a single criterion. This idea is used in [28] and [20] for the first time.

***Theorem 2:*** *(An inner bound on the one-shot capacity region of PP-QWTC)* For any fixed $\epsilon_1 \in (0,1), \epsilon_2 \in (0,1)$ and $\delta_1, \delta_2$ such that $\delta_1 \in (0, \epsilon_1)$ and $\delta_2 \in (0, \epsilon_2)$, there exists a one-shot code for the channel $\mathcal{N}_{X \to YZ}$, if rate pair $(R_1, R_2, 3\epsilon_1 + 2\sqrt{\epsilon_1} + 2\sqrt{\epsilon_2}, 2(\epsilon_1 + \sqrt{\epsilon_1}) + \sqrt{\epsilon_2})$ satisfies the following bounds:

$$R_1 \leq I_H^{\epsilon_1 - \delta_1}(U_1; Y|U_2)_\rho - \tilde{I}_{max}^{\sqrt{\epsilon_2} - \delta_2}(U_1; Z)_\rho - \log\frac{4\epsilon_1}{\delta_1^2} - 2\log\frac{1}{\delta_2}$$

$$R_2 \leq I_H^{\epsilon_1 - \delta_1}(U_2; Y|U_1)_\rho - \tilde{I}_{max}^{\sqrt{\epsilon_2} - \delta_2}(U_2; Z|U_1)_\rho - \log\frac{4\epsilon_1}{\delta_1^2} - 2\log\frac{1}{\delta_2}$$

with respect to state $\rho_{U_1 U_2 YZ} = \sum_{u_2=1}^{|U_2|}\sum_{u_1=1}^{|U_1|} p(u_1, u_2)|u_1\rangle\langle u_1| \otimes |u_2\rangle\langle u_2| \otimes \rho_{YZ}^{u_1 u_2}$.

*Proof*: In Appendix C.

**Remark 2:** The proof of Theorem 2 has two advantages over the proof of Theorem 1: The first is that the proof of Theorem 2 is based on solving a binary hypothesis testing problem against the proof of Theorem 1, which is based on solving a multiple hypothesis testing problem. The second is that in the privacy proof of Theorem 1, Lemma 1 [16] is used. But, in the proof of Theorem 2 the convex split lemma (Lemma 2) can be used.

**Remark 3:** From a comparison between the results of Theorem 1 and Theorem 2, it can be understood that the proof of Theorem 3 does not give the sum-rate $(R_1 + R_2)$. This is because of using the successive decoding technique. This issue should not cause doubts about whether PP-QWTC is a dual for CQ-MA-WTC or not. To solve this doubt, we propose the issue of quantum broadcast channels.

*Quantum broadcast channels*

The quantum broadcast channel (QBC) accepts one user and two or more receivers. In the basic case, the sender (Alice) wishes to transmit three separate messages: $m_1$ is the personal message for the first receiver $Y_1$, $m_2$ is the personal message for the second receiver $Y_2$, and $m_c$ is the common message for both of the receivers.

The basic QBC is illustrated in Figure 3. It should be noted that, for ease of calculation, we removed the security constraint from the problem.

$$\rho_{X_1 X_1' X_1''} \equiv \sum_{x_1} p_{X_1}(x_1)|x_1\rangle\langle x_1|_{X_1} \otimes |x_1\rangle\langle x_1|_{X_1'} \otimes |x_1\rangle\langle x_1|_{X_1''} \tag{6}$$

$$\sigma_{X_2 X_2' X_2''} \equiv \sum_{x_2} p_{X_2}(x_2)|x_2\rangle\langle x_2|_{X_2} \otimes |x_2\rangle\langle x_2|_{X_2'} \otimes |x_2\rangle\langle x_2|_{X_2''} \tag{7}$$

$$\rho_{X_1 X_1'' YZ} \equiv \sum_{x_1} p_{X_1}(x_1)|x_1\rangle\langle x_1|_{X_1} \otimes \rho_{YZ}^{x_1 x_2} \otimes |x_1\rangle\langle x_1|_{X_1''} \tag{8}$$

$$\sigma_{X_2 X_2'' YZ} \equiv \sum_{x_2} p_{X_2}(x_2)|x_2\rangle\langle x_2|_{X_2} \otimes \rho_{YZ}^{x_1 x_2} \otimes |x_2\rangle\langle x_2|_{X_2''} \tag{9}$$

$$\rho_{X_1 X_2 YZ} \equiv \mathcal{N}_{X_1' X_2' \to YZ}\left(\rho_{X_1 X_1'' YZ} \otimes \sigma_{X_2 X_2'' YZ}\right) = \sum_{x_1 x_2} p_{X_1}(x_1)p_{X_2}(x_2)\,|x_1\rangle\langle x_1|_{X_1} \otimes |x_2\rangle\langle x_2|_{X_2} \otimes \rho_{YZ}^{x_1 x_2} \tag{10}$$

$$\frac{1}{|\mathcal{M}_1||\mathcal{M}_2|}\sum_{m_2=1}^{|\mathcal{M}_2|}\sum_{m_1=1}^{|\mathcal{M}_1|}\frac{1}{2}\left\|\mathcal{D}_{Y\to\widehat{M}_1\widehat{M}_2}\left(\rho^{(m_1,k_1),(m_2,k_2)}_{(X_1 X_1'')^{\otimes|\mathcal{M}_1||\mathcal{K}_1|}(X_2 X_2'')^{\otimes|\mathcal{M}_2||\mathcal{K}_2|}YZ}\right) - \left(|m_1\rangle\langle m_1|_{\widehat{M}_1} \otimes |m_2\rangle\langle m_2|_{\widehat{M}_2}\right.\right.$$
$$\left.\left. \otimes \hat{\rho}_{X_1''^{\otimes|\mathcal{M}_1||\mathcal{K}_1|}X_2''^{\otimes|\mathcal{M}_2||\mathcal{K}_2|}Z}\right)\right\|_1 \le \epsilon + 2\delta + 20\delta'^{\frac{1}{8}} \tag{11}$$

where $\hat{\rho}_{X_1''^{\otimes|\mathcal{M}_1||\mathcal{K}_1|}X_2''^{\otimes|\mathcal{M}_2||\mathcal{K}_2|}Z} := \rho^{(m_1,k_1),(m_2,k_2)}_{X_1''^{\otimes|\mathcal{M}_1||\mathcal{K}_1|}X_2''^{\otimes|\mathcal{M}_2||\mathcal{K}_2|}} \otimes \tilde{\rho}_Z$.

$$\mathcal{D}_{Y\to\widehat{M}_1\widehat{M}_2}\left(\rho^{(m_1,k_1),(m_2,k_2)}_{X_1\otimes|\mathcal{M}_1||\mathcal{K}_1|X_2\otimes|\mathcal{M}_2||\mathcal{K}_2|YZ}\right) := \sum_{m_2=1}^{|\mathcal{M}_2|}\sum_{m_1=1}^{|\mathcal{M}_1|} p_{\widehat{M}_1}(m_1)p_{\widehat{M}_1}(m_1)|m_1\rangle\langle m_1|_{\widehat{M}_1} \otimes |m_2\rangle\langle m_2|_{\widehat{M}_2}$$
$$= \sum_{m_2=1}^{|\mathcal{M}_2|}\sum_{m_1=1}^{|\mathcal{M}_1|} Tr\left\{\Lambda^{m_1 m_2}_{X_1\otimes|\mathcal{M}_1||\mathcal{K}_1|X_2\otimes|\mathcal{M}_2||\mathcal{K}_2|Y}\rho^{m_1 m_2 k_1 k_2}_{X_1\otimes|\mathcal{M}_1||\mathcal{K}_1|X_2\otimes|\mathcal{M}_2||\mathcal{K}_2|Y}\right\}|m_1\rangle\langle m_1|_{\widehat{M}_1} \otimes |m_2\rangle\langle m_2|_{\widehat{M}_2} \tag{12}$$

---

The problem of QBC is widely studied in the i.i.d. case in [2-3] and in the one-shot case in [29]. In the following, we want to achieve a one-shot inner bound for QBC with classical messages. Suppose that Alice has not a personal message for the second receiver $Y_2$ ($m_2 = \emptyset \to R_2 = 0$).

The QBC under the one-shot setting is a triple ($\mathcal{X}, \mathcal{N}^{\mathcal{X}\to Y_1 Y_2} \equiv \rho_x^{Y_1 Y_2}, \mathcal{H}^{Y_1} \otimes \mathcal{H}^{Y_2}$), where $\mathcal{X}$ denotes the input alphabet set, and $Y_1, Y_2$ denote the output systems. $\rho_x^{Y_1 Y_2}$ is the system output's quantum state.

**Theorem 3:** *(one-shot inner bound for QBC) Let U be an auxiliary random variable, $p = p_{X|U}(x|u)p_U(u)$ be the code probability function. The one-shot achievable rate consists of all rate pairs $(R_1, R_c)$ such that:*

$$R_1 \le I_H^\epsilon(X; Y_1|U)_\rho - 2 + \log \epsilon$$

$$R_c \le I_H^\epsilon(U; Y_2)_\rho - 2 + \log \epsilon$$

$$R_1 + R_c \le I_H^\epsilon(X; Y_1)_\rho - 2 + \log \epsilon$$

*is achievable, and all information quantities are taken with respect to the following state:*

$$\rho_{UXY_1Y_2} = \sum_{u,x} p_U(u)p_{X|U}(x|u)\,|u\rangle\langle u|_U \otimes |x\rangle\langle x|_X \otimes \rho_x^{Y_1 Y_2} \tag{13}$$

*Proof:* In Appendix D.

Now, consider the extended version of the above theorem:

**Corollary 2:** *(one-shot inner bound for QBC with three personal messages for the first receiver) Let U be an auxiliary random variable, $p = p_U(u)p_{X_1|U}(x_1|u)p_{X_2|UX_1}(x_2|ux_1)$ be the code probability function. The one-shot achievable rate region consists of all rate tuples $(R_1, R_c, R_2)$ in order to sending $(m_1, m_2, m_c)$ such that:*

$$R_1 \le I_H^\epsilon(X_1; Y_1|U)_\rho - 2 + \log \epsilon$$

$$R_2 \le I_H^\epsilon(X_2; Y_1|UX_1)_\rho - 2 + \log \epsilon$$

$$R_c \le I_H^\epsilon(U; Y_2)_\rho - 2 + \log \epsilon$$

$$R_1 + R_2 \le I_H^\epsilon(X_1 X_2; Y_1|U)_\rho - 2 + \log \epsilon$$

$$R_1 + R_c \le I_H^\epsilon(X_1; Y_1)_\rho - 2 + \log \epsilon$$

$$R_2 + R_c \le I_H^\epsilon(X_2; Y_1|X_1)_\rho - 2 + \log \epsilon$$

*is achievable, and all information quantities are taken with respect to the following state:*

$$\rho_{UX_1X_2Y_1Y_2} = \sum_{u,x} p_U(u)p_{X_1|U}(x_1|u)\,p_{X_2|UX_1}(x_2|ux_1)|u\rangle\langle u|_U$$
$$\otimes |x_1\rangle\langle x_1|_{X_1} \otimes |x_2\rangle\langle x_2|_{X_2} \otimes \rho_{X_1 X_2}^{Y_1 Y_2}$$

$$\Lambda_{X_1 \otimes |\mathcal{M}_1||\mathcal{K}_1|X_2 \otimes |\mathcal{M}_2||\mathcal{K}_2|Y}^{(m_1,k_1),(m_2,k_2)}$$

$$:= \left(\sum_{m_2'}\sum_{m_1'}\sum_{k_2'}\sum_{k_1'} \Gamma_{X_1^{|\mathcal{M}_1||\mathcal{K}_1|}X_2^{|\mathcal{M}_2||\mathcal{K}_2|}Y}^{m_1',k_1',m_2',k_2'}\right)^{-\frac{1}{2}} \Gamma_{X_1^{|\mathcal{M}_1||\mathcal{K}_1|}X_2^{|\mathcal{M}_2||\mathcal{K}_2|}Y}^{m_1,k_1,m_2,k_2} \left(\sum_{m_2'}\sum_{m_1'}\sum_{k_2'}\sum_{k_1'} \Gamma_{X_1^{|\mathcal{M}_1||\mathcal{K}_1|}X_2^{|\mathcal{M}_2||\mathcal{K}_2|}Y}^{m_1',k_1',m_2',k_2'}\right)^{-\frac{1}{2}} \quad (14)$$

$$\Gamma_{X_1^{|\mathcal{M}_1||\mathcal{K}_1|}X_2^{|\mathcal{M}_2||\mathcal{K}_2|}Y}^{m_1,k_1,m_2,k_2}$$

$$:= I_{X_1X_2}^{(1,1),(1,1)} \otimes \ldots \otimes I_{X_1X_2}^{(1,1),(1,k_2)} \otimes \ldots \otimes I_{X_1X_2}^{(1,1),(m_2,k_2-1)} \otimes \ldots \otimes I_{X_1X_2}^{(1,k_1),(m_2,k_2)} \otimes \ldots \otimes I_{X_1X_2}^{(m_1,k_1-1),(m_2,k_2)}$$
$$\otimes T_{X_1X_2Y}^{(m_1,k_1),(m_2,k_2)} \otimes I_{X_1X_2}^{(m_1,k_1),(m_2,k_2+1)} \otimes \ldots \otimes I_{X_1X_2}^{(|\mathcal{M}_1|,|\mathcal{K}_1|),(|\mathcal{M}_2|,|\mathcal{K}_2|)} \quad (15)$$

$$Tr\left\{\left(I - \Gamma_{X_1^{|\mathcal{M}_1||\mathcal{K}_1|}X_2^{|\mathcal{M}_2||\mathcal{K}_2|}Y}^{m_1,k_1,m_2,k_2}\right)\rho_{X_1 \otimes |\mathcal{M}_1||\mathcal{K}_1|X_2 \otimes |\mathcal{M}_2||\mathcal{K}_2|Y}^{m_1 m_2 k_1 k_2}\right\} = Tr\left\{(I - T_{X_1X_2Y})\mathcal{N}_{X_1'X_2' \to YZ}\left(\rho_{X_1X_2'} \otimes \sigma_{X_1X_2'}\right)\right\} \quad (16)$$

$$Pr\left((\hat{M}_1, \hat{M}_2) \neq (M_1, M_2)\right)$$
$$\leq (1+c)Tr\left\{\left(I - \Gamma_{X_1^{|\mathcal{M}_1||\mathcal{K}_1|}X_2^{|\mathcal{M}_2||\mathcal{K}_2|}Y}^{m_1,k_1,m_2,k_2}\right)\rho_{X_1 \otimes |\mathcal{M}_1||\mathcal{K}_1|X_2 \otimes |\mathcal{M}_2||\mathcal{K}_2|Y}^{m_1 m_2 k_1 k_2}\right\}$$
$$+ (2+c+c^{-1}) \sum_{m_2' \neq m_2}\sum_{m_1' \neq m_1}\sum_{k_2' \neq k_2}\sum_{k_1' \neq k_1} Tr\left\{\Gamma_{X_1^{|\mathcal{M}_1||\mathcal{K}_1|}X_2^{|\mathcal{M}_2||\mathcal{K}_2|}Y}^{m_1',k_1',m_2',k_2'}\rho_{X_1 \otimes |\mathcal{M}_1||\mathcal{K}_1|X_2 \otimes |\mathcal{M}_2||\mathcal{K}_2|Y}^{m_1 m_2 k_1 k_2}\right\}$$
$$= (1+c)Tr\left\{\left(I - \Gamma_{X_1^{|\mathcal{M}_1||\mathcal{K}_1|}X_2^{|\mathcal{M}_2||\mathcal{K}_2|}Y}^{m_1,k_1,m_2,k_2}\right)\rho_{X_1 \otimes |\mathcal{M}_1||\mathcal{K}_1|X_2 \otimes |\mathcal{M}_2||\mathcal{K}_2|Y}^{m_1 m_2 k_1 k_2}\right\}$$
$$+ (2+c+c^{-1}) \sum_{m_1' \neq m_1}\sum_{k_1' \neq k_1} Tr\left\{\Gamma_{X_1^{|\mathcal{M}_1||\mathcal{K}_1|}X_2^{|\mathcal{M}_2||\mathcal{K}_2|}Y}^{m_1',k_1',m_2,k_2}\rho_{X_1 \otimes |\mathcal{M}_1||\mathcal{K}_1|X_2 \otimes |\mathcal{M}_2||\mathcal{K}_2|Y}^{m_1 m_2 k_1 k_2}\right\}$$
$$+ (2+c+c^{-1}) \sum_{m_2' \neq m_2}\sum_{k_2' \neq k_2} Tr\left\{\Gamma_{X_1^{|\mathcal{M}_1||\mathcal{K}_1|}X_2^{|\mathcal{M}_2||\mathcal{K}_2|}Y}^{m_1,k_1,m_2',k_2'}\rho_{X_1 \otimes |\mathcal{M}_1||\mathcal{K}_1|X_2 \otimes |\mathcal{M}_2||\mathcal{K}_2|Y}^{m_1 m_2 k_1 k_2}\right\}$$
$$+ (2+c+c^{-1}) \sum_{m_2' \neq m_2}\sum_{m_1' \neq m_1}\sum_{k_2' \neq k_2}\sum_{k_1' \neq k_1} Tr\left\{\Gamma_{X_1^{|\mathcal{M}_1||\mathcal{K}_1|}X_2^{|\mathcal{M}_2||\mathcal{K}_2|}Y}^{m_1',k_1',m_2',k_2'}\rho_{X_1 \otimes |\mathcal{M}_1||\mathcal{K}_1|X_2 \otimes |\mathcal{M}_2||\mathcal{K}_2|Y}^{m_1 m_2 k_1 k_2}\right\}$$
$$= (1+c)Tr\left\{(I - T_{X_1X_2Y})\mathcal{N}_{X_1'X_2' \to YZ}\left(\rho_{X_1X_2'} \otimes \sigma_{X_1X_2'}\right)\right\}$$
$$+ (2+c+c^{-1})(|\mathcal{M}_1||\mathcal{K}_1| - 1)Tr\left\{(I - T_{X_1X_2Y})\mathcal{N}_{X_1'X_2' \to YZ}\left(\rho_{X_1} \otimes \rho_{X_2'} \otimes \sigma_{X_1X_2'}\right)\right\}$$
$$+ (2+c+c^{-1})(|\mathcal{M}_2||\mathcal{K}_2| - 1)Tr\left\{(I - T_{X_1X_2Y})\mathcal{N}_{X_1'X_2' \to YZ}\left(\rho_{X_1X_2'} \otimes \sigma_{X_1} \otimes \sigma_{X_2'}\right)\right\}$$
$$+ (2+c+c^{-1})(|\mathcal{M}_1||\mathcal{K}_1| - 1)(|\mathcal{M}_2||\mathcal{K}_2|$$
$$- 1)Tr\left\{(I - T_{X_1X_2Y})\mathcal{N}_{X_1'X_2' \to YZ}\left(\rho_{X_1} \otimes \rho_{X_2'} \otimes \sigma_{X_1} \otimes \sigma_{X_2'}\right)\right\} \quad (17)$$

---

*Proof*: The proof follows the extended version of Theorem 3's proof.

The channel described in Corollary 2 will be converted to the channel described in Theorem 2 (PP-QWTC) without secrecy constraint by choosing $m_c = \emptyset$. Set $R_c = 0$ in Corollary 2:

$$R_1 \leq I_H^\epsilon(X_1; Y_1)_\rho - 2 + \log \epsilon$$
$$R_2 \leq I_H^\epsilon(X_2; Y_1|X_1)_\rho - 2 + \log \epsilon$$
$$R_1 + R_2 \leq I_H^\epsilon(X_1X_2; Y_1)_\rho - 2 + \log \epsilon$$
$$R_1 \leq I_H^\epsilon(X_1; Y_1)_\rho - 2 + \log \epsilon$$
$$R_2 \leq I_H^\epsilon(X_2; Y_1|X_1)_\rho - 2 + \log \epsilon$$

where the above last two rates are redundant. Then, we have the following region:

$$R_1 \leq I_H^\epsilon(X_1; Y_1)_\rho - 2 + \log \epsilon$$
$$R_2 \leq I_H^\epsilon(X_2; Y_1|X_1)_\rho - 2 + \log \epsilon \quad (18)$$
$$R_1 + R_2 \leq I_H^\epsilon(X_1X_2; Y_1)_\rho - 2 + \log \epsilon$$

Consider the results of Theorem 2 without the leaked information terms. By choosing $\delta_2 = \epsilon_2$, we have:

$$R_1 \leq I_H^\epsilon(X_1; Y_1|X_2)_\rho - 2 + \log \epsilon_1$$
$$R_2 \leq I_H^\epsilon(X_2; Y_1|X_1)_\rho - 2 + \log \epsilon_1 \quad (19)$$

Comparing (18), (19), and (33), the argument stated in Remark 3 is proved. Also, the region (18) is a *near-optimal* achievable rate region compared to Corollary 1. As it can be understood from a comparison between the results of Corollary 1, Theorem 2, and Corollary 2 by considering (31), this idea can be proved that converting the CQ-MA-WTC to PP-QWTC can be a helpful approach to bypass the bottlenecks connected to the multiple hypothesis testing problem (Theorem 1) and the smoothing bottlenecks of quantum information theory (Corollary 1 and Theorem 1).

*Asymptotic analysis*

In this subsection, we want to evaluate our secrecy rate region in the asymptotic i.i.d. case (asymptotic limit of many uses of a memoryless channel). Consider PP-QWTC $(x(u_1, u_2) \to \rho_x^{YZ})$. The capacity region of the channel can be expressed as follows:

$$\mathcal{C}_\infty(\mathcal{N}) \coloneqq \lim_{\epsilon_1,\epsilon_2 \to 0} \lim_{n \to \infty} \frac{1}{n} \mathcal{C}^{\epsilon_1,\epsilon_2}(\mathcal{N}^{\otimes n}) \quad (20)$$

where $\mathcal{C}^{\epsilon_1,\epsilon_2}(\mathcal{N}^{\otimes n}) \equiv \max_{p(u_1,u_2)} \mathcal{R}^{\epsilon_1,\epsilon_2}(\mathcal{N}^{\otimes n})$.

Let $\mathcal{R}(\mathcal{N})$ be the set of the maximum rate pairs $(R'_1, R'_2)$,

$$\mathcal{R}(\mathcal{N}) = \begin{cases} R'_1 \leq I(U_1; Y|U_2)_\rho - I(U_1; Z)_\rho \\ R'_2 \leq I(U_2; Y|U_1)_\rho - I(U_2; Z|U_1)_\rho \end{cases} \quad (21)$$

Then the capacity region $\mathcal{C}_\infty(\mathcal{N})$ is the union over $n$ uses of the channel $\mathcal{N}$:

$$\mathcal{C}_\infty(\mathcal{N}) \coloneqq \max_{p(u_1,u_2)} \frac{1}{n} \bigcup_{n=1}^{\infty} \mathcal{R}(\mathcal{N}^{\otimes n}) \quad (22)$$

Our aim is to prove the expression above. Consider both of single rates. Applying Fact 2 (and its conditional version) we have,

$$R_1 \geq I_H^{\epsilon_1 - \delta_1}(U_1; Y|U_2)_\rho - I_{max}^{\sqrt{\epsilon_2} - \delta_2 - \gamma}(U_2; Z)_\rho - \log \frac{4\epsilon_1}{\delta_1^2}$$
$$- 2\log \frac{1}{\delta_2} - \log \frac{3}{\gamma^2}$$

$$R_2 \geq I_H^{\epsilon_1 - \delta_1}(U_2; Y|U_1)_\rho - I_{max}^{\sqrt{\epsilon_2} - \delta_2 - \gamma}(U_2; Z|U_1)_\rho - \log \frac{4\epsilon_1}{\delta_1^2}$$
$$- 2\log \frac{1}{\delta_2} - \log \frac{3}{\gamma^2}$$

To prove the achievability, consider the one-shot lower bounds presented in Theorem 2, and apply quantum AEP [30] for the conditional smooth hypothesis testing, and max-mutual information. From Theorem 2, for $r$ uses of the channel $\mathcal{N}$, the following lower bound $\mathcal{C}^{\epsilon_1,\epsilon_2}(\mathcal{N}^{\otimes n})$ can be obtained:

$$\bigcup_{n=1}^{r} \mathcal{R}(\mathcal{N}^{\otimes n}) \subseteq \mathcal{C}^{\epsilon_1,\epsilon_2}(\mathcal{N}^{\otimes r})$$

where $\mathcal{R}(\mathcal{N}^{\otimes n})$ is the set of all rate pairs $(R'_1, R'_2)$ satisfying:

$$R'_1 \leq I_H^{\epsilon_1 - \delta_1}(U_1^n; Y^{\otimes n}|U_2^n)_\rho - I_{max}^{\sqrt{\epsilon_2} - \delta_2 - \gamma}(U_1^n; Z^{\otimes n})_\rho$$
$$- \log \frac{4\epsilon_1}{\delta_1^2} - 2\log \frac{1}{\delta_2} - \log \frac{3}{\gamma^2} \quad (23)$$

$$R'_2 \leq I_H^{\epsilon_1 - \delta_1}(U_2^n; Y^{\otimes n}|U_1^n)_\rho$$
$$- I_{max}^{\sqrt{\epsilon_2} - \delta_2 - \gamma}(U_2^n; Z^{\otimes n}|U_1^n)_\rho$$
$$- \log \frac{4\epsilon_1}{\delta_1^2} - 2\log \frac{1}{\delta_2} - \log \frac{3}{\gamma^2} \quad (24)$$

We can assume that the sequences of the random variables are generated in an i.i.d. fashion according to their distributions. This is due to the fact that the region above is basically a lower bound on the capacity region. This empowers us to make use of quantum AEP as described below. From Fact 3, we have,

$$\lim_{\epsilon_1 \to 0} \lim_{r \to \infty} \frac{1}{r} I_H^{\epsilon_1 - \delta_1}(U_1^r; Y^{\otimes r}|U_2^r)_{\rho^{\otimes r}} = I(U_1; Y|U_2)_\rho \quad (25)$$

$$\lim_{\epsilon_1 \to 0} \lim_{r \to \infty} \frac{1}{r} I_H^{\epsilon_1 - \delta_1}(U_2^r; Y^{\otimes r}|U_1^r)_{\rho^{\otimes r}} = I(U_2; Y|U_1)_\rho \quad (26)$$

Also, using Fact 4, we have the following:

$$\lim_{\epsilon_2 \to 0} \lim_{r \to \infty} \frac{1}{r} I_{max}^{\sqrt{\epsilon_2} - \delta_2 - \gamma}(U_1^r; Z^{\otimes r})_{\rho^{\otimes r}} = I(U_1; Z|U_2)_\rho \quad (27)$$

$$\lim_{\epsilon_2 \to 0} \lim_{r \to \infty} \frac{1}{r} I_{max}^{\sqrt{\epsilon_2} - \delta_2 - \gamma}(U_2^r; Z^{\otimes r}|U_1^r)_{\rho^{\otimes r}} \quad (28)$$
$$= I(U_2; Z|U_1)_\rho$$

Putting (25), (26), (27), and (28) into (23), and (24) gives (21):

$$\mathcal{R}(\mathcal{N}^{\otimes n}) \subseteq \lim_{\epsilon_1,\epsilon_2 \to 0} \lim_{r \to \infty} \frac{1}{r} \mathcal{C}^{\epsilon_1,\epsilon_2}(\mathcal{N}^{\otimes r})$$

Given the argument above and using (20), and (22) completes the proof.

V. DISCUSSION

In this paper, we studied the problem of secure communication over a CQ-MA-WTC using three techniques: 1- Sen's joint typicality lemma. 2-simultaneous position-based decoding and, 3-successive position-based decoding. The first and the second decoding techniques use a newly introduced smooth technique [16] to analyze the privacy, while the third technique uses convex splitting [19]. We realized that the simultaneous position-based decoder tends to a multiple hypothesis testing problem which is unsolvable in the general case. We introduced a new channel (PP-QWTC) which can be considered as a dual for CQ-MA-WTC. Also, this channel can be derivate from the quantum broadcast channel. The results show that the PP-QWTC has a near-optimal achievable rate region to CQ-MA-WTC.

APPENDIX

*Appendix A: (Proof of Corollary 1)*

As mentioned, the proof has two steps: Reliable decoding and secure decoding. To these ends, consider two junk variables $k_i; i \in \{1,2\}$ for each of users $m_i, i \in \{1,2\}$. These junk variables are used to make two doubly indexed codebooks $\{x_1(m_1, k_1)\}_{m_1 \in \mathcal{M}_1, k_1 \in \mathcal{K}_1}$ and $\{x_2(m_2, k_2)\}_{m_2 \in \mathcal{M}_2, k_2 \in \mathcal{K}_2}$. Bob should be able to detect the pair messages $(m_1, m_2)$, and the junk variables $k_1$, and $k_2$ with high probability.

Using Definition 10 (Sen's inner bound for QMAC), we have the following relation:

$$\mathcal{R}_{priv-QMA-WTC} = \mathcal{R}_{Sen} - \mathcal{R}_{leaked}$$

with decoding error at most $49\sqrt{\epsilon}$, and privacy leakage at most $20\delta'^{\frac{1}{8}}$ (Lemma 1). Also, $\mathcal{R}_{Sen}$ refers to Sen's inner bound for QMAC (Definition 9), and $\mathcal{R}_{leaked}$ refers to the leaked information from senders to Eve.

From Lemma 1, we have the following:

$$R_{1-leaked} \leq I_{max}^{\delta'-\epsilon'}(X_1:Z)_\rho + \log\frac{3}{\epsilon'^3} - \frac{1}{4}\log\delta'$$

$$R_{2-leaked} \leq I_{max}^{\delta'-\epsilon'}(X_2:ZX_1)_\rho + \log\frac{3}{\epsilon'^3} - \frac{1}{4}\log\delta' + \mathcal{O}(1)$$

This completes the proof.

***Appendix B:*** *(Proof of Theorem 1)*

Both of the messages are uniformly distributed on their sets. The receiver has to be able to decode both messages with negligible error probability. Before communication begins, Alice (A) and Bob (B) share randomness with Charlie (C), and wiretapper (Z). Let $\rho_{X_1 X_1' X_1''}$ (6) and $\sigma_{X_2 X_2' X_2''}$ (7) be shared-randomness between (A,C,Z) and shared-randomness between (B,C,Z), respectively. Alice has $X_1'$ system, Bob has $X_2'$ system, and Charlie has $(X_1, X_2)$ system, and wiretapper has $(X_1'', X_2'')$ system. Let $\rho_{X_1 X_1'' YZ}$ (8) and $\sigma_{X_2 X_2'' YZ}$ (9) denote the state resulting from sending $X_1'$ and $X_2'$ over the channel, respectively. Then the overall controlling state of the channel is as stated in (10).

*Sketch of the coding scheme:* For each of the messages $(m_i), i \in \{1,2\}$, there exist local keys $k_i \in [1:|\mathcal{K}_i|], i \in \{1,2\}$ as uniform randomness for randomizing Eve's knowledge about the sent messages. These local keys are not accessible to Charlie or Eve. Before the communication begins, assume that Alice, Charlie, and Eve share $|\mathcal{M}_1||\mathcal{K}_1|$ copies of the state in (6) and Bob, Charlie and Eve share $|\mathcal{M}_2||\mathcal{K}_2|$ copies of the state in (7):

$$\rho_{X_1^{|\mathcal{M}_1||\mathcal{K}_1|} X_1'^{|\mathcal{M}_1||\mathcal{K}_1|} X_1''^{|\mathcal{M}_1||\mathcal{K}_1|}} = \rho_{X_1 X_1' X_1''}^{\otimes |\mathcal{M}_1||\mathcal{K}_1|}$$

$$\sigma_{X_2^{|\mathcal{M}_2||\mathcal{K}_2|} X_2'^{|\mathcal{M}_2||\mathcal{K}_2|} X_2''^{|\mathcal{M}_2||\mathcal{K}_2|}} = \sigma_{X_2 X_2' X_2''}^{\otimes |\mathcal{M}_2||\mathcal{K}_2|}$$

To send the pair messages $m_1$, and $m_2$, Alice and Bob pick $k_1 \in [1:|\mathcal{K}_1|]$ and $k_2 \in [1:|\mathcal{K}_2|]$, respectively, and uniformly at random. They send $(m_1, k_1)$-th system $X_1'$ and $(m_2, k_2)$-th system $X_2'$ through the channel $\mathcal{N}_{X_1' X_2' \to YZ}$.

There exists a simultaneous decoder for communication over a CQ-MA-WTC with the upper bound on the average error probability, as stated in (11). As it can be understood from (11), the security criterion is merged into the reliability criterion [20]. The simultaneous position-based decoder can be constructed as stated in (12), where,

$$\Lambda_{X_1 \otimes |\mathcal{M}_1||\mathcal{K}_1| X_2 \otimes |\mathcal{M}_2||\mathcal{K}_2|_Y}^{m_1 m_2} = \sum_{k_2=1}^{|\mathcal{K}_2|} \sum_{k_1=1}^{|\mathcal{K}_1|} \Lambda_{X_1 \otimes |\mathcal{M}_1||\mathcal{K}_1| X_2 \otimes |\mathcal{M}_2||\mathcal{K}_2|_Y}^{(m_1,k_1),(m_2,k_2)}$$

Now, we consider the error term. Charlie constructs her position-based decoder to decode $m_1$, $m_2$, $k_1$, and $k_2$. Let $\Lambda_{X_1 \otimes |\mathcal{M}_1||\mathcal{K}_1| X_2 \otimes |\mathcal{M}_2||\mathcal{K}_2|_Y}^{(m_1,k_1),(m_2,k_2)}$ denotes the POVM:

$$Tr\left\{\left(I_{X_1^{|\mathcal{M}_1||\mathcal{K}_1|} X_2^{|\mathcal{M}_2||\mathcal{K}_2|}_Y} - \Lambda_{X_1 \otimes |\mathcal{M}_1||\mathcal{K}_1| X_2 \otimes |\mathcal{M}_2||\mathcal{K}_2|_Y}^{(m_1,k_1),(m_2,k_2)}\right) \rho_{X_1 \otimes |\mathcal{M}_1||\mathcal{K}_1| X_2 \otimes |\mathcal{M}_2||\mathcal{K}_2|_Y}^{m_1 m_2 k_1 k_2}\right\} \leq \epsilon$$

where $\Lambda_{X_1 \otimes |\mathcal{M}_1||\mathcal{K}_1| X_2 \otimes |\mathcal{M}_2||\mathcal{K}_2|_Y}^{(m_1,k_1),(m_2,k_2)}$ is expressed in (14), and for $m_i \in [1:|\mathcal{M}_i|]$ and $k_i \in [1:|\mathcal{K}_i|]$, $\Gamma_{X_1^{|\mathcal{M}_1||\mathcal{K}_1|} X_2^{|\mathcal{M}_2||\mathcal{K}_2|}_Y}^{m_1,k_1,m_2,k_2}$ is expressed in (15), in which $T_{X_1 X_2 Y}^{(m_1,k_1,m_2,k_2)}$ is a test operator used to discriminate between hypotheses $\rho_{X_1 X_2 Y}, \rho_{X_1 X_2} \otimes \rho_Y, \rho_{X_1} \otimes \rho_{X_2 Y}$ and $\rho_{X_1} \otimes \rho_{X_2} \otimes \rho_Y$ with an error of $\epsilon$. Note that, this hypothesis testing problem is equal to discriminating between hypotheses $\mathcal{N}_{X_1' X_2' \to YZ}(\rho_{X_1 X_2'} \otimes \sigma_{X_1 X_2'})$, $\mathcal{N}_{X_1' X_2' \to YZ}(\rho_{X_1 X_2'} \otimes \sigma_{X_1} \otimes \sigma_{X_2'})$, $\mathcal{N}_{X_1' X_2' \to YZ}(\rho_{X_1} \otimes \rho_{X_2'} \otimes \sigma_{X_1 X_2'})$ and $\mathcal{N}_{X_1' X_2' \to YZ}(\rho_{X_1} \otimes \rho_{X_2'} \otimes \sigma_{X_1} \otimes \sigma_{X_2'})$. Therefore, if Charlie checks for message pair $(m_1, m_2)$ when message pair $(m_1, m_2)$ is actually transmitted, then the probability of incorrectly decoding is as stated in (16).

Similarly, other kinds of error probabilities can be considered as:

- If Charlie checks for message pair $(m_1, m_2)$ when message pair $(m_1', m_2)$ is indeed transmitted, then the probability of incorrectly decoding is:

$$Tr\left\{\left(I - \Gamma_{X_1^{|\mathcal{M}_1||\mathcal{K}_1|} X_2^{|\mathcal{M}_2||\mathcal{K}_2|}_Y}^{m_1',k_1,m_2,k_2}\right) \rho_{X_1 \otimes |\mathcal{M}_1||\mathcal{K}_1| X_2 \otimes |\mathcal{M}_2||\mathcal{K}_2|_Y}^{m_1 m_2 k_1 k_2}\right\}$$
$$= Tr\{(I - T_{X_1 X_2 Y})\mathcal{N}_{X_1' X_2' \to YZ}(\rho_{X_1} \otimes \rho_{X_2'} \otimes \sigma_{X_1 X_2'})\} \quad (29)$$

- If Charlie checks for message pair $(m_1, m_2)$ when message pair $(m_1, m_2')$ is indeed transmitted, then the probability of incorrectly decoding is:

$$Tr\left\{\left(I - \Gamma_{X_1^{|\mathcal{M}_1||\mathcal{K}_1|} X_2^{|\mathcal{M}_2||\mathcal{K}_2|}_Y}^{m_1,k_1,m_2',k_2}\right) \rho_{X_1 \otimes |\mathcal{M}_1||\mathcal{K}_1| X_2 \otimes |\mathcal{M}_2||\mathcal{K}_2|_Y}^{m_1 m_2 k_1 k_2}\right\}$$
$$= Tr\{(I - T_{X_1 X_2 Y})\mathcal{N}_{X_1' X_2' \to YZ}(\rho_{X_1 X_2'} \otimes \sigma_{X_1} \otimes \sigma_{X_2'})\} \quad (30)$$

- If Charlie checks for message pair $(m_1, m_2)$ when message pair $(m_1', m_2')$ is indeed transmitted, then the probability of incorrectly decoding is:

$$Tr\left\{\left(I - \Gamma_{X_1^{|\mathcal{M}_1||\mathcal{K}_1|} X_2^{|\mathcal{M}_2||\mathcal{K}_2|}_Y}^{m_1',k_1,m_2',k_2}\right) \rho_{X_1 \otimes |\mathcal{M}_1||\mathcal{K}_1| X_2 \otimes |\mathcal{M}_2||\mathcal{K}_2|_Y}^{m_1 m_2 k_1 k_2}\right\}$$
$$= Tr\{(I - T_{X_1 X_2 Y})\mathcal{N}_{X_1' X_2' \to YZ}(\rho_{X_1} \otimes \rho_{X_2'} \otimes \sigma_{X_1} \otimes \sigma_{X_2'})\} \quad (31)$$

Due to the code construction, the error probability under the position-based coding scheme is the same for each message pair $(m_1, m_2)$:

$$Pr\left((\widehat{M}_1, \widehat{M}_2) \neq (M_1, M_2)\right)$$
$$= Tr\left\{\left(I - \Lambda_{X_1 \otimes |\mathcal{M}_1||\mathcal{K}_1| X_2 \otimes |\mathcal{M}_2||\mathcal{K}_2|_Y}^{(m_1,k_1),(m_2,k_2)}\right) \rho_{X_1 \otimes |\mathcal{M}_1||\mathcal{K}_1| X_2 \otimes |\mathcal{M}_2||\mathcal{K}_2|_Y}^{m_1 m_2 k_1 k_2}\right\}$$

Applying Lemma 3 with $S = \Gamma_{X_1^{|\mathcal{M}_1||\mathcal{K}_1|} X_2^{|\mathcal{M}_2||\mathcal{K}_2|}_Y}^{m_1,k_1,m_2,k_2}$ and $T = \sum_{m_2' \neq m_2} \sum_{m_1' \neq m_1} \sum_{k_2' \neq k_2} \sum_{k_1' \neq k_1} \Gamma_{X_1^{|\mathcal{M}_1||\mathcal{K}_1|} X_2^{|\mathcal{M}_2||\mathcal{K}_2|}_Y}^{m_1',k_1',m_2',k_2'}$, we have a

chain of equalities and inequalities as stated in (17). Note that, we used (29)-(31).

*Multiple quantum hypothesis testing:* As mentioned before, the problem of the existence of a simultaneous decoder for a general QMAC (more than two users) remained an open problem in the i.i.d. case. In [17], the authors presented a helpful discussion about the multiple quantum hypothesis testing and its relation with QMACs. In summary, the problem of multiple hypothesis testing is an open problem too. There are two possible hypothesis testing schemes: Symmetric and asymmetric. *Chernoff distance* from symmetric hypothesis testing gives a lower bound on the randomness-assisted error exponent [31]; In contrast, the application of the results from asymmetric hypothesis testing leads to a lower bound on the one-shot randomness-assisted capacity (for QMAC without secrecy constraint) and in turn on the second-order coding rate for randomness-assisted communication.

In other words, from [7], we know that there exists a general simultaneous decoder to decode more than two messages simultaneously in the case of commutative version of outputs, and from [17], we know that the multiple hypothesis testing problem can be solved if the composite alternative hypothesis forms a commutative set of operators. This means that, for a test operator $T$, a finite set of positive semi-definite operators $\theta \equiv \{\theta_i : 1 \leq i \leq r\}$, for which $supp(\rho) \subseteq supp(\theta_i)$ and $\min_i D(\rho \| \theta_i) > 0$, there are two hypotheses, and we have:

$$Tr\{(I-T)\rho\} \leq \epsilon \quad (32)$$

$$-\log_2 Tr\{T\theta_i\} \geq \left[\min_i D(\rho \| \theta_i)\right] - \delta \quad (33)$$

where $\delta$ is a positive integer.

The last inequality holds when the set $\theta$ forms a commutative set of operators. More information can be found in [17].

With these explanations, we use asymmetric hypothesis testing for our problem. Note that we want to decode two messages simultaneously. Consider the upper bound on error probability in (17). Then, we rewrite that as follows:

$$Pr\left((\widehat{M}_1, \widehat{M}_2) \neq (M_1, M_2)\right)$$
$$\leq (1+c)Tr\{(I-T)\mu\}$$
$$+ (2+c+c^{-1})Tr\{T(\theta_1 + \theta_2 + \theta_3)\}$$

where,

$$\mu = \mathcal{N}_{X_1'X_2' \to YZ}\left(\rho_{X_1X_2'} \otimes \sigma_{X_1X_2'}\right)$$
$$\theta_1 = \mathcal{N}_{X_1'X_2' \to YZ}\left(\rho_{X_1} \otimes \rho_{X_2'} \otimes \sigma_{X_1X_2'}\right)$$
$$\theta_2 = \mathcal{N}_{X_1'X_2' \to YZ}\left(\rho_{X_1X_2'} \otimes \sigma_{X_1} \otimes \sigma_{X_2'}\right)$$
$$\theta_3 = \mathcal{N}_{X_1'X_2' \to YZ}\left(\rho_{X_1} \otimes \rho_{X_2'} \otimes \sigma_{X_1} \otimes \sigma_{X_2'}\right)$$

This is called asymmetric hypothesis testing, which tries to minimize all other probabilities subject to a constraint on the error probability $Tr\{(I-T)\mu\}$. Note that we consider all three hypotheses $(\theta_1 + \theta_2 + \theta_3)$ as a unique composite alternative hypothesis.

We can say for such a sequence of test operators, as stated in (32), and (33), the above multiple hypothesis testing problem can be solved as:

$$Pr\left((\widehat{M}_1, \widehat{M}_2) \neq (M_1, M_2)\right)$$
$$\leq (1+c)Tr\{(I-T)\mu\}$$
$$+ (2+c+c^{-1})Tr\{T(\theta_1 + \theta_2 + \theta_3)\}$$
$$= (1+c)\epsilon$$
$$+ (2+c+c^{-1})\{|\mathcal{K}_1|2^{R_1 - D_H^\epsilon(\mu\|\theta_1)}$$
$$+ |\mathcal{K}_2|2^{R_2 - D_H^\epsilon(\mu\|\theta_2)}$$
$$+ |\mathcal{K}_1||\mathcal{K}_2|2^{R_1+R_2 - D_H^\epsilon(\mu\|\theta_3)}\}$$
$$= (1+c)\epsilon$$
$$+ (2+c+c^{-1})\{|\mathcal{K}_1|2^{R_1 - I_H^\epsilon(X_1 : X_2Y)}$$
$$+ |\mathcal{K}_2|2^{R_2 - I_H^\epsilon(X_2 : X_1Y)}$$
$$+ |\mathcal{K}_1||\mathcal{K}_2|2^{R_1+R_2 - I_H^\epsilon(X_1X_2 : Y)}\}$$

Let $|\mathcal{K}_1| = 2^{\hat{R}_1}$ and $|\mathcal{K}_2| = 2^{\hat{R}_2}$. Then, by setting the above term equal to $\epsilon$, with a straightforward simplification, we have:

$$R_1 + \hat{R}_1 = I_H^\epsilon(X_1 : X_2Y) + \log_2\left(\frac{\epsilon - (1+c)\epsilon}{2+c+c^{-1}}\right)$$

$$R_2 + \hat{R}_2 = I_H^\epsilon(X_2 : X_1Y) + \log_2\left(\frac{\epsilon - (1+c)\epsilon}{2+c+c^{-1}}\right)$$

$$R_1 + \hat{R}_1 + R_2 + \hat{R}_2 = I_H^\epsilon(X_1X_2 : Y) + \log_2\left(\frac{\epsilon - (1+c)\epsilon}{2+c+c^{-1}}\right)$$

The global maximum of the above expression with respect to $c$ occurs at $c = \frac{\delta}{\epsilon}$:

$$R_1 + \hat{R}_1 = I_H^\epsilon(X_1 : X_2Y) - \log_2\left(\frac{4\epsilon}{\delta^2}\right) \quad (34)$$

$$R_2 + \hat{R}_2 = I_H^\epsilon(X_2 : X_1Y) - \log_2\left(\frac{4\epsilon}{\delta^2}\right) \quad (35)$$

$$R_1 + \hat{R}_1 + R_2 + \hat{R}_2 = I_H^\epsilon(X_1X_2 : Y) - \log_2\left(\frac{4\epsilon}{\delta^2}\right) \quad (36)$$

and for such a $c$, we have:

$$Pr\left((\widehat{M}_1, \widehat{M}_2) \neq (M_1, M_2)\right) \leq \epsilon + 2\delta \quad (37)$$

Now, we turn our attention to the secrecy criterion. Using Lemma 1, we have:

$$\hat{R}_1 \leq I_{max}^{\delta' - \epsilon'}(X_1 : Z)_\rho + \log\frac{3}{\epsilon'^3} - \frac{1}{4}\log\delta' \quad (38)$$

$$\hat{R}_2 \leq I_{max}^{\delta' - \epsilon'}(X_2 : ZX_1)_\rho + \log\frac{3}{\epsilon'^3} - \frac{1}{4}\log\delta' + \mathcal{O}(1) \quad (39)$$

Substituting (38) and (39) in (34)-(36) completes the proof.

*Appendix C: (Proof of Theorem 2)*

The proof uses two successive position-based decoders. The first decoder tries to decode the first message $m_1$, and the second decoder tries to decode the second message $m_2$ given the true decoded $m_1$. This means that if the first decoder fails, the second decoder fails too. This decoding order can be shown as $m_1 \to m_2$.

Constructing the first position-based decoder is the same as that presented in [20]. To decode $m_2$, Bob performs his second position-based decoder conditioned on $U_1$, which works for all $u_1 \in \mathcal{U}_1$. It should be noted that, the feeding state of the second decoder differs from the main state of the channel.

Alice, Bob, and Eve are allowed to pre-share some quantum state as randomness. Also, Alice has access to two sources of uniform junk randomness $k_i; i \in \{1,2\}$. The pre-shared randomness is as follows:

$$\rho_{X_1 X_1'(AX_2 X_2')^{\otimes |\mathcal{M}_2||\mathcal{K}_2|}}^{\otimes |\mathcal{M}_1||\mathcal{K}_1|}$$
$$:= \left[ \sum_{x_1} p(x_1) |x_1\rangle\langle x_1|_{X_1} \right.$$
$$\otimes |x_1\rangle\langle x_1|_{X_1'} \left( \sum_{x_2} p(x_2) |x_2\rangle\langle x_2|_{X_2} \right.$$
$$\left. \left. \otimes |x_2\rangle\langle x_2|_{X_2'} \right)^{\otimes |\mathcal{M}_2||\mathcal{K}_2|} \right]^{\otimes |\mathcal{M}_1||\mathcal{K}_1|} \quad (40)$$

The arguments connected to the decoding process for $m_1$ are listed as follows:

- The probability of error for decoding $m_1$:
$$p_{e_1} = p\{\widehat{M}_1 \neq M_1\}$$
$$:= \frac{1}{|\mathcal{M}_1|} \sum_{m_1=1}^{|\mathcal{M}_1|} \frac{1}{2} \left\| \mathcal{D}_{YX_1 \to \widehat{M}_1}^{m_1} \left( \rho_{X_1 \otimes |\mathcal{M}_1||\mathcal{K}_1|Y}^{(m_1,k_1),(m_2,k_2)} \right) \right.$$
$$\left. - |m_1\rangle\langle m_1|_{\widehat{M}_1} \otimes \hat{\rho}_Z \right\|_1 \leq \epsilon_1 + \sqrt{\epsilon_2} \quad (41)$$

where $\mathcal{D}_{YX_1 \to \widehat{M}_1}^{m_1} \left( \rho_{X_1 \otimes |\mathcal{M}_1||\mathcal{K}_1|Y}^{(m_1,k_1),(m_2,k_2)} \right)$ is decoding map for $m_1$:

$$\mathcal{D}_{YX_1 \to \widehat{M}_1}^{m_1} \left( \rho_{X_1 \otimes |\mathcal{M}_1||\mathcal{K}_1|Y}^{(m_1,k_1),(m_2,k_2)} \right)$$
$$:= \sum_{k_1=1}^{|\mathcal{K}_1|} \sum_{m_1=1}^{|\mathcal{M}_1|} Tr \left\{ \Lambda_{X_1|\mathcal{M}_1||\mathcal{K}_1|Y}^{m_1,k_1} \rho_{X_1 \otimes |\mathcal{M}_1||\mathcal{K}_1|Y}^{(m_1,k_1),(m_2,k_2)} \right\}$$
$$\otimes \frac{\sqrt{\Lambda_{X_1|\mathcal{M}_1||\mathcal{K}_1|Y}^{m_1,k_1}} \rho_{X_1 \otimes |\mathcal{M}_1||\mathcal{K}_1|Y}^{(m_1,k_1),(m_2,k_2)} \sqrt{\Lambda_{X_1|\mathcal{M}_1||\mathcal{K}_1|Y}^{m_1,k_1}}}{Tr \left\{ \Lambda_{X_1|\mathcal{M}_1||\mathcal{K}_1|Y}^{m_1,k_1} \rho_{X_1 \otimes |\mathcal{M}_1||\mathcal{K}_1|Y}^{(m_1,k_1),(m_2,k_2)} \right\}}$$

- $\Lambda_{X_1|\mathcal{M}_1||\mathcal{K}_1|Y}^{m_1,k_1}$ is a pretty good measurement (POVM) for $m_1 \in [1:|\mathcal{M}_1|]$:

$$\Lambda_{X_1|\mathcal{M}_1||\mathcal{K}_1|Y}^{m_1,k_1}$$
$$:= \left( \sum_{k_1'=1}^{|\mathcal{K}_1|} \sum_{m_1'=1}^{|\mathcal{M}_1|} \Gamma_{X_1|\mathcal{M}_1||\mathcal{K}_1|Y}^{m_1',k_1'} \right)^{-1/2} \Gamma_{X_1|\mathcal{M}_1||\mathcal{K}_1|Y}^{m_1,k_1} \left( \sum_{k_1'=1}^{|\mathcal{K}_1|} \sum_{m_1'=1}^{|\mathcal{M}_1|} \Gamma_{X_1|\mathcal{M}_1||\mathcal{K}_1|Y}^{m_1',k_1'} \right)^{-1/2}$$

where $\Gamma_{X_1|\mathcal{M}_1||\mathcal{K}_1|Y}^{m_1,k_1}$ is the element of the first POVM:

$$\Gamma_{X_1|\mathcal{M}_1||\mathcal{K}_1|Y}^{m_1,k_1} := I_{X_1}^{(1,1)} \otimes \ldots \otimes I_{X_1}^{(1,|\mathcal{K}_1|)} \otimes \ldots \otimes \tau_{X_1Y}^{m_1,k_1} \otimes \ldots$$
$$\otimes I_{X_1}^{(|\mathcal{M}_1|,|\mathcal{K}_1|)}$$

and $\tau_{X_1Y}^{m_1,k_1}$ is a test operator in order to discriminate between two hypotheses $\rho_{X_1Y}$, and $\rho_{X_1} \otimes \rho_Y$. Also, it is obvious that to decode $m_1$, it does not matter for the second position-based decoder, which copy is selected by Alice among $|\mathcal{M}_2||\mathcal{K}_2|$ copies.

- We face a hypothesis testing problem. Null hypothesis is $\rho_{X_1Y}$ and alternative hypothesis is $\rho_{X_1} \otimes \rho_Y$. Therefore the probability of success in guessing null and alternative hypotheses are $Tr\{\tau_{X_1Y} \rho_{X_1Y}\}$ and $Tr\{(I_{X_1Y} - \tau_{X_1Y})(\rho_{X_1} \otimes \rho_Y)\}$.

The rest of the decoding process for $m_1$ is analogous to [20]. Therefore, we have:

$$R_1 \leq I_H^{\epsilon_1-\delta_1}(U_1;Y)_\rho - \tilde{I}_{max}^{\sqrt{\epsilon_2}-\delta_2}(U_1;Z)_\rho - \log \frac{4\epsilon_1}{\delta_1^2}$$
$$- 2\log \frac{1}{\delta_2} \quad (42)$$

Now, we turn our attention to decoding the second message. As mentioned before, the channel state changes after the first measurement. There is a detailed discussion in [32].

Let $\sigma_{X_1X_1'(X_2X_2')^{\otimes|\mathcal{M}_2||\mathcal{K}_2|}YZ}^{(m_1,k_1),(m_2,k_2)}$ denote the disturbed state after applying the first measurement (POVM):

$$\sigma_{X_1X_1'(X_2X_2')^{\otimes|\mathcal{M}_2||\mathcal{K}_2|}YZ}^{(m_1,k_1),(m_2,k_2)}$$
$$:= \sum_{x_1} p_{X_1}(x_1) |x_1\rangle\langle x_1|_{X_1} \otimes |x_1\rangle\langle x_1|_{X_1'}$$
$$\otimes \sigma_{X_2X_2'}^{x_1,(m_1,k_1)} \otimes \ldots \otimes \sigma_{X_2X_2'YZ}^{x_1,(m_1,k_1),(m_2,k_2)} \otimes \ldots$$
$$\otimes \sigma_{X_2X_2'}^{x_1,(m_1,k_1),(|\mathcal{M}_2|,|\mathcal{K}_2|)}$$

Also, Bob's second POVM is as follows:

$$\Lambda_{X_1X_2|\mathcal{M}_2||\mathcal{K}_2|Y}^{m_2,k_2} := \left( \sum_{k_2'=1}^{|\mathcal{K}_2|} \sum_{m_2'=1}^{|\mathcal{M}_2|} \lambda_{X_1X_2|\mathcal{M}_2||\mathcal{K}_2|Y}^{m_2',k_2'} \right)^{-1/2} \lambda_{X_1X_2|\mathcal{M}_2||\mathcal{K}_2|Y}^{m_2,k_2}$$
$$\left( \sum_{k_2'=1}^{|\mathcal{K}_2|} \sum_{m_2'=1}^{|\mathcal{M}_2|} \lambda_{X_1X_2|\mathcal{M}_2||\mathcal{K}_2|Y}^{m_2',k_2'} \right)^{-1/2}$$

$\lambda_{X_1X_2|\mathcal{M}_2||\mathcal{K}_2|Y}^{m_2,k_2}$ is the element of the second POVM:

$$\lambda_{X_1X_2|\mathcal{M}_2||\mathcal{K}_2|Y}^{m_2,k_2} := |x_1\rangle\langle x_1|_{X_1} \otimes I_{X_2}^{(1,1)} \otimes \ldots \otimes I_{X_2}^{(1,|\mathcal{K}_2|)} \otimes \ldots \otimes \theta_{X_2Y}^{m_2,k_2}$$
$$\otimes \ldots \otimes I_{X_2}^{(|\mathcal{M}_2|,|\mathcal{K}_2|)}$$

$\theta_{X_2Y}^{m_2,k_2}$ is a binary test operator to discriminate between two hypotheses $\sigma_{X_2Y}^{x_1}$ and $\sigma_{X_2}^{x_1} \otimes \sigma_Y^{x_1}$ with an error of $\epsilon_1 - \delta_1$; i.e.,

$$Tr\{\theta_{X_2Y}\sigma_{X_2Y}^{x_1}\} \geq 1 - (\epsilon_1 - \delta_1) \quad \epsilon_1 \in (0,1), \quad \delta_1 \in (0,\epsilon_1)$$

In other words, Bob has to be able to discriminate between the following states:

$$\sum_{x_1} p_{X_1}(x_1)|x_1\rangle\langle x_1|_{X_1} \otimes \sigma_{X_2Y}^{x_1}$$

$$\sum_{x_1} p_{X_1}(x_1)|x_1\rangle\langle x_1|_{X_1} \otimes \sigma_{X_2}^{x_1} \otimes \sigma_Y^{x_1}$$

Similar to what mentioned in [20], and [27], we have the following rate:

$$R_2 \leq I_H^{\epsilon_1-\delta_1}(U_2;Y|U_1)_\rho - \tilde{I}_{max}^{\sqrt{\epsilon_2}-\delta_2}(U_2;Z|U_1)_\rho - \log\frac{4\epsilon_1}{\delta_1^2} - 2\log\frac{1}{\delta_2} \tag{43}$$

The probability of error for $m_2$ is as follows:

$$\begin{aligned}p_{e_2} &= p\{\widehat{M}_2 \neq M_2\} \\ &:= \frac{1}{|\mathcal{M}_2|}\sum_{m_2=1}^{|\mathcal{M}_2|}\frac{1}{2}\left\|\mathcal{D}_{\widehat{M}_1YX_2\to\widehat{M}_2}^{m_1}\left(\sigma_{X_1X_1'(X_2X_2')^{\otimes|\mathcal{M}_2||\mathcal{K}_2|}YZ}^{(m_1,k_1),(m_2,k_2)}\right)|m_2\rangle\langle m_2|_{\widehat{M}_2} \\ &\otimes \hat{\sigma}_{X_1'X_2'^{\otimes|\mathcal{M}_2||\mathcal{K}_2|}Z}\right\|_1 \leq 2(\epsilon_1 + \sqrt{\epsilon_2}) + \sqrt{\epsilon_1'} \tag{44}\end{aligned}$$

Also, the error probability exponents stated in (41) and (44) are proved. See [20, 27].

This process can be repeated for another decoding order. In other words, we can first decode $m_2$, and then decode $m_1$ ($m_2 \to m_1$). Then taking the intersection of the regions resulting from both orders, we give:

$$R_1 \leq I_H^{\epsilon_1-\delta_1}(U_1;Y|U_2)_\rho - \tilde{I}_{max}^{\sqrt{\epsilon_2}-\delta_2}(U_1;Z)_\rho - \log\frac{4\epsilon_1}{\delta_1^2} - 2\log\frac{1}{\delta_2}$$

$$R_2 \leq I_H^{\epsilon_1-\delta_1}(U_2;Y|U_1)_\rho - \tilde{I}_{max}^{\sqrt{\epsilon_2}-\delta_2}(U_2;Z|U_1)_\rho - \log\frac{4\epsilon_1}{\delta_1^2} - 2\log\frac{1}{\delta_2}$$

This completes the proof.

**Appendix D**: *(Proof of Theorem 3)*

The proof uses superposition coding. Assume that the first receiver $Y_1$ has a better reception signal than the second receiver $Y_2$. In this setting, Alice is able to encode a further message superimposed on top of the common message. Using the successive decoding can be helpful.

*Codebook generation*: Randomly and independently generate $2^{R_c}$ sequence $u(m_c)$ according to the distribution $p_U(u)$. For each sequence $u(m_c)$, randomly and conditionally independently generate $2^{R_1}$ sequence $x(m_1, m_c)$ according to the distribution $p_{X|U}(x|u(m_c))$. The $Y_1$'s state can be calculated by tracing out $Y_2$ from (13):

$$\rho_{UXY_1} = \sum_{u,x} p_U(u)p_{X|U}(x|u)|u\rangle\langle u|_U \otimes |x\rangle\langle x|_X \otimes \rho_x^{Y_1}$$

Similar to what mentioned for Theorem 2, we construct the POVM for the first receiver as:

$$\Lambda_{m_1,m_c} := \left(\sum_{m_c'=1}^{|\mathcal{M}_c|}\sum_{m_1'=1}^{|\mathcal{M}_1|}\Gamma_{m_1',m_c'}\right)^{-1/2} \Gamma_{m_1,m_c} \left(\sum_{m_c'=1}^{|\mathcal{M}_c|}\sum_{m_1'=1}^{|\mathcal{M}_1|}\Gamma_{m_1',m_c'}\right)^{-1/2}$$

Also, the POVM for the second receiver can be constructed as follows:

$$\Lambda_{m_c} := \left(\sum_{m_c'=1}^{|\mathcal{M}_c|}\lambda_{m_c'}\right)^{-1/2} \lambda_{m_c} \left(\sum_{m_c'=1}^{|\mathcal{M}_c|}\lambda_{m_c'}\right)^{-1/2}$$

Consider the probability of error for $m_1$:

$$\begin{aligned}p_{e_1} &= p\{(\widehat{M}_1, \widehat{M}_c) \neq (M_1, M_c)\} \\ &:= \frac{1}{|\mathcal{M}_1||\mathcal{M}_c|}\sum_{m_c}\sum_{m_1} Tr\left\{(I - \Lambda_{m_1,m_c})\rho_{x(m_1,m_c)}^{Y_1}\right\}\end{aligned}$$

and for $m_c$:

$$p_{e_2} = p\{\widehat{M}_c \neq M_c\} := \frac{1}{|\mathcal{M}_c|}\sum_{m_c} Tr\left\{(I - \Lambda_{m_c})\rho_{x(m_c)}^{Y_2}\right\}$$

By a straightforward calculation analogous to [3] for i.i.d. case and in [29] (to calculate one-shot Marton inner bound for QBC), the above error probability exponents can be calculated as follows:

$$p_{e_1} + p_{e_2} \leq 2^{-I_H^\epsilon(X;Y_1|U)_\rho - 2 + \log\epsilon} + 2^{-I_H^\epsilon(U;Y_2)_\rho - 2 + \log\epsilon} + 2^{-I_H^\epsilon(X;Y_1)_\rho - 2 + \log\epsilon} + \mathcal{O}(\epsilon)$$

This completes the proof.